\documentclass[12pt]{article}
\newcommand{\bq}{\begin{equation}}
\newcommand{\eq}{\end{equation}}
\newcommand{\ba}{\begin{eqnarray}}
\newcommand{\ea}{\end{eqnarray}}

\newcommand{\nl }{ \nonumber  }

\newcommand{\ul}{\underline}
\newcommand{\p}{\partial}
\newcommand{\pu}{\p_\tau}
\newcommand{\pJ}{\p_J}
\newcommand{\pK}{\p_K}
\newcommand{\pj}{\p_j}
\newcommand{\pk}{\p_k}
\newcommand{\h}{\hspace{1cm}}
\newcommand{\s}{\sigma}

\newcommand{\us}{\underline\sigma}

\begin{document}
\vspace*{.5cm}
{\bf\begin{center}
  NULL STRINGS AND MEMBRANES IN DEMIANSKI-NEWMAN BACKGROUND
\footnote{Work supported in part by the National Science Foundation
of Bulgaria under contract $\phi-620/1996$} \vspace*{.5cm}\\ P.
Bozhilov \footnote {E-mail: bojilov@thsun1.jinr.ru}, B. Dimitrov
\\ \it Bogoliubov
Laboratory of Theoretical Physics, \\ JINR, 141980 Dubna, Russia
\end{center}}

We consider null bosonic $p$-branes moving in curved space-times.
Some exact solutions of the classical equations of motion and of the
constraints for the null string and the null membrane in
Demianski-Newman background are found.


\section{\bf Introduction}
\hspace{1cm} The null $p$-branes are the zero tension limit of the
usual $p$-branes, the 1-brane being a string. This relationship
between them may be regarded as a generalization of the
correspondence between massless and massive particles. On the other
hand, the characteristic scale of string theory is given by the
string tension $T=(2\pi\alpha ')^{-1}$. Then one may view the high
energy limit of string theory as a {\it zero tension limit}, since
only the energy measured in string units, $E/T^{1/2}$, is relevant.

The investigations in the domain of classical and quantum string
propagation in curved space-times are relevant for the quantum
gravity as well as for the understanding of the cosmic string models
in cosmology. As we have already mentioned, strings are characterized
by an energy scale $T^{1/2}$ and the length of the string scales like
$1/ T^{1/2}$. The gravitational field provides another length scale,
the curvature radius of the space-time $R_c$. For a string moving in
a gravitational field an appropriate parameter is the dimensionless
constant $C=R_c T^{1/2}$. Large values of $C$ imply weak
gravitational field. One may reach large values of $C$ by letting $T
\to {\infty}$. In this limit the string shrinks to a point. In the
opposite limit, small values of $C$, one encounters {\it strong
gravitational fields} and it is appropriate to consider $T \to 0$,
i.e. {\it null} or {\it tensionless} strings.

The motion of classical null bosonic strings in curved backgrounds
have been considered recently in
\cite{RZ95,K96,DL96,DL97,PP97,GKKP98,KKP98}. The dynamics of null
$p$-branes living in Friedmann-Robertson-Walker space-time with flat
space-like section (${\it k}=0$) have been investigated in
\cite{RZ96}.

In this letter we describe the classical evolution of the null
strings and membranes in Demianski-Newman background \cite{DN66}. In
Sec. 2 we begin with the general case of tensionless $p$-branes
moving in $D$-dimensional curved space-times. In Sec. 3 we obtain
some exact solutions of the equations of motion and of the
constraints for the null string $(p=1)$ and the null membrane $(p=2)$
in the above mentioned background. Sec. 4 is devoted to our
concluding remarks.

\section{\bf Null $p$-Branes in Curved Background}
\hspace{1cm} To begin with, let us write down one of the possible
actions for the null bosonic $p$-brane living in a $D$-dimensional
curved space-time with metric tensor $g_{\mu\nu}(x)$: \ba\label{a}
S=\int d^{p+1}\xi {\cal L} \h,\h {\cal L}=V^JV^K\pJ x^\mu\pK x^\nu
g_{\mu\nu}(x),\h\h
\\ \nl
\pJ=\p/\p\xi^J \h, \h \xi^J=(\xi^0,\xi^j)=(\tau,\s^j),\h\h
\\ \nl
J,K=0,1,...,p \h,\h j,k=1,...,p \h,\h \mu,\nu=0,1,...,D-1. \ea It is
an obvious generalization of the flat space-time action given in
\cite{NPCQG94}. One easily verifies that if $x^\mu(\xi)$,
$g_{\mu\nu}(\xi)$ are world-volume scalars and $V^J(\xi)$ is a
world-volume contravariant vector density of weight $q=1/2$, then
${\cal L}(\xi)$ is a scalar density of weight $q=1$. As a
consequence, the variation of the action (\ref{a}) is \ba\nl
\delta_{\varepsilon}S=\int d^{p+1}\xi\p_J\bigl ( \varepsilon^J {\cal
L}\bigr ) \ea and thus this action is invariant under world-volume
reparametrizations if suitable boundary conditions are assumed.

Varying (\ref{a}) with respect to $x^{\nu}$ and $V^J$, one obtains
the following equations of motion : \ba\nl \pJ\Bigl (V^J V^K \pK
x^{\lambda}\Bigr ) + \Gamma^{\lambda}_{\mu\nu} V^J V^K \pJ x^{\mu}\pK
x^{\nu} = 0 ,\\ \nl V^J \pJ x^{\mu}\pK x^{\nu} g_{\mu\nu}(x) = 0 ,
\ea where $\Gamma^{\lambda}_{\mu\nu}$ is the connection compatible
with the metric $g_{\mu\nu}(x)$: \ba\nl
\Gamma^{\lambda}_{\mu\nu}=\frac{1}{2}g^{\lambda\rho}\bigl(
\p_{\mu}g_{\nu\rho}+\p_{\nu}g_{\mu\rho}-\p_{\rho}g_{\mu\nu}\bigr) .
\ea

Let us rewrite for further convenience the Lagrangian density
(\ref{a}) in the form ($\pu=\p/\p\tau, \pj=\p/\p\s^j$): \ba\label{L}
{\cal L}=\frac{1}{4\mu^0} g_{\mu\nu}(x)\bigl (\pu-\mu^j\pj\bigr
)x^\mu \bigl (\pu-\mu^k\pk\bigr )x^\nu , \ea where the connection
between $V^J$ and $(\mu^0,\mu^j)$ is given by \ba\nl
V^J=\bigl(V^0,V^j\bigr)=\Biggl(-\frac{1}{2\sqrt{\mu^0}},
\frac{\mu^j}{2\sqrt{\mu^0}}\Biggr) . \ea Now the Euler-Lagrange
equation for $x^\nu$ takes the form: \ba\label{eqx} \pu\Biggl
[\frac{1}{2\mu^0}\bigl (\pu-\mu^k\pk\bigr )x^{\lambda}\Biggr ]
-\pj\Biggl [\frac{\mu^{j}}{2\mu^{0}}\bigl (\pu-\mu^k\pk\bigr
)x^{\lambda} \Biggr ] \\ \nl +
\frac{1}{2\mu^0}\Gamma^{\lambda}_{\mu\nu} \bigl (\pu-\mu^j\pj\bigr
)x^\mu \bigl (\pu-\mu^k\pk\bigr )x^\nu = 0 . \ea The equations of
motion for the Lagrange multipliers $\mu^{0}$ and $\mu^{j}$ which
follow from (\ref{L}) give the constraints : \ba\label{Tx1}
g_{\mu\nu}(x)\bigl (\pu-\mu^j\pj\bigr )x^\mu \bigl (\pu-\mu^k\pk\bigr
)x^\nu = 0 , \\ \label{Tx2} g_{\mu\nu}(x)\bigl (\pu-\mu^k\pk\bigr
)x^\mu \pj x^\nu = 0 . \ea We note (and it is easy to check) that in
the gauge $\mu^j = const$, \ba\nl
x^{\nu}(\tau,\s^k)=x^{\nu}(\mu^{k}\tau+\s^k) \ea is a nontrivial
solution of the equations of motion (\ref{eqx}) and of the
constraints (\ref{Tx1}), (\ref{Tx2}), depending on $D$ arbitrary
functions of $p$ variables for the null $p$-brane moving in arbitrary
$D$-dimensional gravity background.

In terms of $x^\nu$ and the conjugated momentum $p_\nu$ the
constraints (\ref{Tx1}) can be written as: \ba\nl
T_0=g^{\mu\nu}(x)p_\mu p_\nu = 0 \h,\h T_j=p_\nu\pj x^\nu = 0 . \ea
Then one can show that the Hamiltonian which corresponds to the
Lagrangian density (\ref{L}) is a linear combination of $T_0$ and
$T_j$ \ba\nl H=\int d^p\sigma\bigl (\mu^0 T_0+\mu^j T_j \bigr ) . \ea
Computation of the Poisson brackets between $T_0$ and $T_j$ results
in the following constraint algebra \ba\nl \{T_0(\ul
\sigma_1),T_0(\ul \sigma_2)\}&=&0,
\\ \label {CA}
\{T_0(\ul \sigma_1),T_{j}(\ul \sigma_2)\}&=& [T_0(\ul \sigma_1) +
T_0(\ul \sigma_2)] \p_j \delta^p (\ul \sigma_1 - \ul \sigma_2) ,
\\ \nl
\{T_{j}(\ul \sigma_1),T_{k}(\ul \sigma_2)\}&=&
[\delta_{j}^{l}T_{k}(\ul \sigma_1) + \delta_{k}^{l}T_{j}(\ul
\sigma_2)]\p_l\delta^p(\ul \sigma_1-\ul \sigma_2) ,\\ \nl
\us=(\s^1,...,\s^p) . \ea Comparing the equalities (\ref{CA}) with
the zero tension limit of the bosonic $p$-brane flat space-time
constraint algebra \cite{BN90} we see that they coincide. However
this is not the case with the Hamiltonian equations of motion of
course. These are: \ba\label{hem} \bigl (\pu-\mu^k\pk\bigr )x^\nu =
2\mu^0 g^{\nu\lambda}(x)p_{\lambda} , \\ \nl \pu p_{\rho}-\pk (\mu^k
p_{\rho}) + \mu^0 \p_{\rho}g^{\nu\lambda}(x)p_{\nu}p_{\lambda} = 0 .
\ea

General solutions of the equations (\ref{hem}) can be easily found
only in flat space-times. For example, the general solution for the
null string case in an arbitrary gauge is \cite{B991}: \ba \nl
x^\nu(\tau,\s^{1})&=&g^\nu(v)-2\int\limits_{}^{\s^{1}}
\frac{\mu^0(s)}{[\mu^{1}(s)]^2}ds f^\nu(v) ,\\ \nl
p_\nu(\tau,\s^{1})&=&f_\nu(v)/\mu^{1}(\s^{1}) ,\ea where $g^\nu(v)$,
$f_\nu(v)$ are arbitrary functions of the variable \ba\nl v = \tau +
\int\limits^{\s^{1}} \frac{ds}{\mu^{1}(s)}. \ea In curved space-times
one can try to find backgrounds in which the system of partial
differential equations (\ref{hem}) essentially simplifies. An example
of this type are the conformally flat spaces, in which the equation
of motion for $p_{\nu}$ (in the gauge $\mu^{j}= constants$) reduces
to $\bigl (\pu-\mu^j\pj\bigr )p_\nu = 0$ if we take into account the
constraint $T_0=0$, and has as a general solution
$p_{\nu}(\tau,\sigma^{k}) = p_{\nu}(\mu^{k}\tau + \sigma^{k})$.

\section{\bf Null Strings and Membranes \\ in Demianski-Newman Space-Time}
\hspace{1cm} In this section we will work in the gauge $\mu^0, \mu^j
= constants$, in which the Euler-Lagrange equations (\ref{eqx}) have
the form : \ba\label{eqxf} \bigl(\pu-\mu^j\pj\bigr)\bigl
(\pu-\mu^k\pk\bigr )x^{\lambda} + \Gamma^{\lambda}_{\mu\nu} \bigl
(\pu-\mu^j\pj\bigr )x^\mu \bigl (\pu-\mu^k\pk\bigr )x^\nu = 0 . \ea
We are going to look for solutions of the equations of motion
(\ref{eqxf}) and constraints (\ref{Tx1}), (\ref{Tx2}), for the null
string $(p=1)$ and the null membrane $(p=2)$ moving in
Demianski-Newman background \cite{DN66}. The metric for this
space-time is of the following type: \ba\nl
ds^2&=&g_{00}(dx^0)^2+2g_{01}dx^0 dx^1+2g_{03}dx^0 dx^3+2g_{13}dx^1
dx^3+g_{22}(dx^2)^2+ g_{33}(dx^3)^2 .\ea It can be put in a form in
which the manifest expressions for $g_{\mu\nu}$ are given by the
equalities (all other components are zero) \cite{KSMH80}:  \ba\nl
g_{00}&=&- \exp(+2U) \h,\h g_{11}=
\Biggl(1-\frac{a^2\sin^2\theta}{{\cal R}^2}\Biggr)\exp(-2U),
\\ \label{dnm}
g_{22}&=& \left({\cal R}^2 - a^2\sin^2\theta\right)\exp(-2U),\\ \nl
g_{33}&=& {\cal R}^2\sin^2\theta\exp(-2U)
-\Biggl[\frac{2(Mr+l)a\sin^2\theta + 2{\cal R}^{2}l\cos\theta}{{\cal
R}^2 - a^2\sin^2\theta}\Biggr]^2\exp(+2U) ,
\\ \nl
g_{03}&=&- \frac{2(Mr+l)a\sin^2\theta + 2{\cal
R}^{2}l\cos\theta}{{\cal R}^2 - a^2\sin^2\theta}\exp(+2U), \ea where
\ba\nl \exp(\pm 2U) = \Biggl[1-2\frac{Mr + l\left(a\cos\theta +
l\right)}{r^2 + \left(a\cos\theta + l\right)^2}\Biggr]^{\pm 1},\h
{\cal R}^2 = r^2 - 2Mr + a^2 - l^2 , \ea $M$ is a mass parameter, $a$
is an angular momentum per unit mass and $l$ is the NUT parameter.
The metric (\ref{dnm}) is a stationary axisymmetric metric. It
belongs to the vacuum solutions of the Einstein field equations,
which are of type D under Petrov's classification. The Kerr and NUT
space-times are particular cases of the considered metric and can be
obtained by putting $l=0$ or $a=0$ in (\ref{dnm}). The case $l=0,a=0$
obviously corresponds to the Schwarzschild solution.

Starting with the string case, we introduce the ansatz \ba\nl
x^0(\tau,\sigma^1)&=&C^0f(z^1)+t(\tau) ,
\\ \label{az}
x^1(\tau,\sigma^1)&=&r(\tau) \h,\h x^2(\tau,\sigma^1)=\theta(\tau) ,
\\ \nl
x^3(\tau,\sigma^1)&=&C^3f(z^1)+\varphi(\tau) ,
\\ \nl
z^1&=&\mu^1 \tau + \sigma^1 , \ea where $f(z^1)$ is an arbitrary
function of $z^1$ and $C^0$, $C^3$ are constants. The substitution of
(\ref{az}) in (\ref{Tx1}), (\ref{Tx2}) and (\ref{eqxf}), for the
given metric, allows us to obtain a solution of the equations for
$\dot t\equiv \p t/\p\tau$, $\dot\varphi\equiv\p\varphi/\p\tau$ and
constraints (\ref{Tx2}) in the form $(C_1=const)$: \ba\nl \dot
t(\tau)=-C_1\left(C^0 g_{03}+C^3 g_{33}\right)\exp(-{\cal H}) ,
\\ \label{s032}
\dot\varphi (\tau)=+C_1\left(C^0 g_{00}+C^3 g_{03}\right)\exp(-{\cal
H}) ,
\\ \nl
{\cal H} = \int\Bigl(g^{00}dg_{00}+2g^{03}dg_{03}+g^{33}dg_{33}\Bigr)
. \ea The condition for compatibility of this solution with the other
equations and constraints (\ref{Tx1}) reads: \ba\nl \exp({\cal
H})\equiv h=g_{00}g_{33}-g_{03}^2 . \ea Then one can show that the
solutions for $\dot r=\p r/\p\tau$ and $\dot \theta=\p\theta/\p\tau$
are expressed by the equalities: \ba\nl \dot
r^2&=&-g^{11}\Biggl[C_1^2\frac{G}{h}+g^{22}\left(g_{22}^2
\dot\theta^2\right)\Biggr] ,\hspace{.3cm} G = (C^0)^2 g_{00}+2C^0 C^3
g_{03}+(C^3)^2 g_{33},
\\ \label{t.2}
g_{22}^2\dot\theta^2&=&C_2(r) + C_1^2\int\limits^{\theta}d\theta
h^{-2} \Biggl[g_{22}G\frac{\p h}{\p\theta}-h\frac{\p
g_{22}G}{\p\theta}\Biggr] . \ea

In obtaining (\ref{t.2}), we have used that for the metric
(\ref{dnm}) the following condition is fulfilled: $\p_2
\bigl(g_{22}/g_{11}\bigr) = 0$.

In the particular case when $x^2=\theta=\theta_0=const$, one can
integrate the equations of motion and constraints completely to
obtain an exact solution for the null string in the gravity
background (\ref{dnm}). This solution is given by (\ref{az}), where

\ba\nl t-t_0&=&\pm\int\limits_{r_0}^{r}dr\Biggl[\frac{\left(C^0 + C^3
{\cal A}_0\right){\cal A}_0}{{\cal R}^2\sin^2\theta_0}\exp(+2U_0) -
C^3\exp(-2U_0)\Biggr]W^{-1/2} ,
\\ \label{GSC}
\varphi - \varphi_{0}&=&\mp\int\limits_{r_0}^{r}dr \frac{C^0 + C^3
{\cal A}_0}{{\cal R}^2\sin^2\theta_0}\exp(+2U_0)W^{-1/2} ,
\\ \nl
C_1 (\tau -\tau_{0})&=& \pm\int\limits_{r_0}^{r}drW^{-1/2} ,\\ \nl
W&=&\left({\cal R}^2-a^2\sin^2\theta_0\right)^{-1}
\Biggl[\left(C^3\right)^2{\cal R}^2 -  \frac{\left(C^0 + C^3 {\cal
A}_0\right)^2}{\sin^2\theta_0}\exp(+4U_0)\Biggr], \\ \nl {\cal
A}_0&=&\frac{2(Mr+l)a\sin^2\theta_0 + 2{\cal
R}^{2}l\cos\theta_0}{{\cal R}^2 - a^2\sin^2\theta_0},\h
U_0=U_{|\theta=\theta_0},
\\ \nl
t_0, r_0, \varphi_{0}, \tau_{0} &-& constants. \ea

For the null membrane, we use the ansatz \ba\nl x^0(\tau,\us)&=&C^0
F[w(z^1,z^2)]+t(\tau) ,
\\ \label{maz}
x^1(\tau,\us)&=&r(\tau) \h,\h x^2(\tau,\us)=\theta(\tau) ,
\\ \nl
x^3(\tau,\us)&=&C^3 F[w(z^1,z^2)]+\varphi(\tau) ,
\\ \nl
z^j&=&\mu^j \tau + \sigma^j ,\h j=1,2, \ea where the arbitrary
function $F$ depends on $z^j$ only by the one variable $w$. Then one
proceeds as in the null string case to obtain an exact solution for
the null membrane in the Demianski-Newman background. It is given by
(\ref{maz}) with $\theta=\theta_0=const$ and $t,\varphi$ and $\tau$
taken from (\ref{GSC}).

\section{\bf Conclusions}
\hspace{1cm} In this letter we perform some investigations on the
classical dynamics of the null bosonic branes in curved space-times.
In the second section we give the action, show that it is
reparametrization invariant and write down the equations of motion
and constraints in an arbitrary gauge. Then we construct the
corresponding Hamiltonian and compute the constraint algebra. It
coincides with the flat space-time algebra. The Hamiltonian equations
of motion are also given. In the third section we consider the
dynamics of the null strings and membranes in the Demianski-Newman
background. Some exact solutions of the equations of motion and of
the constraints are found. In particular, our null string solution
(\ref{az}), (\ref{GSC}) generalizes the one obtained in \cite{PP97}
for the Kerr space-time.

\vspace*{.5cm}


\end{document}